\documentclass[hyper,12pt,paper]{article}
\pdfoutput=1

\usepackage{jheppub}
\usepackage{amsmath}
\usepackage{graphicx}
\usepackage{cancel}
\usepackage[dvipsnames]{xcolor}
\usepackage{color}
\usepackage{mathrsfs}
\usepackage{amsfonts}
\usepackage{dsfont}
\usepackage{color}
\usepackage{tikz}
\usepackage{ulem}
\usepackage{cancel}

\usepackage{subcaption}
\usepackage{float}
\usepackage{afterpage}
\allowdisplaybreaks

\def\({\left(}
\def\){\right)}
\def\[{\left[}
\def\]{\right]}

\newcommand{\be}{\begin{equation}}
\newcommand{\ee}{\end{equation}}

\newcommand{\ban}[1]{\begin{align}#1\end{align}}

\newcommand{\Tr}{\text{Tr}}
\newcommand{\ket}[1]{\left| #1\right>}

\title{Handlebody phases and the polyhedrality of the holographic entropy cone}

\author{Donald Marolf, Massimiliano Rota, and Jason Wien}

\affiliation{Department of Physics, University of California, Santa Barbara, CA 93106, USA}

\emailAdd{marolf@physics.ucsb.edu}
\emailAdd{mrota@physics.ucsb.edu}
\emailAdd{jswien@physics.ucsb.edu}

\abstract{The notion of a holographic entropy cone has recently been introduced and it has been proven that this cone is polyhedral. However, the original definition was fully geometric and did not strictly require a holographic duality. We introduce a new definition of the cone, insisting that the geometries used for its construction should be dual to states of a CFT. As a result,  the polyhedrality of this holographic cone does not immediately follow. A numerical evaluation of the Euclidean action for the geometries that realize extremal rays of the original cone indicates that these are subdominant bulk phases of natural path integrals. The result challenges the expectation that such geometries are in fact dual to CFT states.
}

\makeatletter
\def\@fpheader{\relax}
\makeatother

\begin{document}
\maketitle


\section{Introduction}
\label{•}

Since the original proposal by Ryu and Takayanagi \cite{rt1} for the calculation of entanglement entropy in holographic field theories, the inequalities implied by the ``RT formula'' have been a subject of active investigations. It was already noted in \cite{rt2} that \textit{subadditivity} ($\mathcal{SA}$) of the von Neumann entropy is satisfied by the prescription, while the celebrated proof \cite{Headrick:2007aa} of the more restrictive \textit{strong subadditivity} ($\mathcal{SSA}$) served as a further important check. The saturation of $\mathcal{SA}$ was later associated to a phase transition of mutual information in \cite{Headrick:2010aa} and the saturation of the Araki-Lieb inequality $(\mathcal{AL})$ to the entanglement plateaux phenomenon \cite{Hubeny:2013aa}.

While all the previous inequalities were already known from quantum mechanics (they are in fact satisfied by any quantum state), the first purely ``holographic'' inequality was found in \cite{Hayden:2011aa} and dubbed \textit{monogamy of mutual information} $(\mathcal{MMI})$.\footnote{See \cite{Rangamani:2015aa,Hosur:2015aa,Rota:2015aa} for more details.} This inequality is not satisfied by all quantum states and provides a new constraint on the family of states which admit a classical geometric dual. It is then interesting to ask whether the RT formula implies additional constraints for a fixed (but arbitrary) number of regions in a holographic CFT. The systematic study of this problem was initiated in \citep{Bao}, which proved that no new inequalities exist for three or four regions and found new ones for five.\footnote{In addition, \citep{Bao} also proved a new family of inequalities for any odd number of regions. The inequality for 3 regions coincides with $\mathcal{MMI}$.}

In principle, for a fixed number of regions in the dual field theory, the set of entropy inequalities which are satisfied by holographic states could be rather complicated. For example, there could be an infinite number of them or they could be nonlinear. Instead, an important result of \citep{Bao} was the proof that not only the number of inequalities implied by the RT prescription is finite for any number of regions, but all inequalities are in fact linear and with integer coefficients.

For an arbitrary system of inequalities it is useful to consider a geometric representation where the set of solutions is a region bounded by the corresponding hypersurfaces. For the inequalities implied by RT this is then a rational polyhedron in the space of entropies, in fact a cone, which has an equivalent description in terms of its extremal rays. Any ray inside the cone (or on its boundary) can be obtained from a (conical) linear combination of them.

Motivated by holography one then tries to construct geometries that realize the extremal rays and to show that such geometries are in fact dual to some field theory state. This would prove that the region of entropy space corresponding to states of holographic CFTs with classical bulk duals is a rational polyhedron. The present work focuses on this last statement and argues that the proof of \citep{Bao} was not conclusive. This naturally raises the question of whether there are any additional entropic constraints on holographic states not implied by Ryu-Takayanagi in the manner of \citep{Bao}.

We will focus on three dimensional bulk geometries with a moment of time symmetry, such that the time slice is in general a disjoint union of Riemann surfaces with boundaries. This can be obtained by slicing the full three dimensional Euclidean solution, which is a handlebody obtained by filling in the boundary Riemann surface \cite{Krasnov1,Skenderis:2009aa}. Different fillings will correspond to different phases of the bulk geometry \cite{MBW1}. One can then evaluate the Euclidean action for the different phases to determine which one is the dominant saddle. Via the AdS/CFT dictionary this will give an approximation to the corresponding state in the dual field theory. On the other hand, if the bulk solution is not the dominant saddle, one cannot immediately conclude that this geometry is dual to a field theory state.

This is the argument which we will employ in the following to challenge the conclusion that the entropy cone of holographic CFTs is polyhedral. We will evaluate numerically the Euclidean action for the particular geometries corresponding to some of the extremal rays of the cone in the $N=3$ and $N=4$ cases and provide evidence suggesting that these are never dominant. While it remains possible that such geometries may nevertheless be related to CFT states by more complicated constructions, such an analysis is beyond the scope of this work.

The structure of the paper is as follows. In \S\ref{sec:cones} we review the definition of \citep{Bao} and introduce a new one to make a clear distinction between a notion of the cone which is purely geometric, and one that is instead more specific to holographic states dual to classical geometries.  In \S\ref{sec:rays} we evaluate the Euclidean actions for the geometries that realize some of the extremal rays of the $N=3,4$ cones. We summarize our conclusions in \S\ref{sec:discussion} and comment on some open questions.

\section{Entropy cones}
\label{sec:cones}

In this section we first review the definition of the \textit{quantum entropy cone} for arbitrary quantum systems, and the definition given in \citep{Bao} for the holographic context. We warn the reader that we will change terminology from the one used in \citep{Bao}. What we will call the \textit{metric entropy cone} is precisely the construction of \citep{Bao}, instead we reserve the name \textit{RT (or HRT) cone} for a different object that we will define in \S\ref{subsec:holographiccone}. The conceptual distinction between the two constructions, and the question of whether they do or do not coincide, is the main motivation for the calculation that we will present in \S\ref{sec:rays}.

\subsection{The Quantum Entropy Cone ($\mathcal{QC}_N$)}
\label{subsec:quantumcone}

Consider a multipartite quantum system associated to a Hilbert space $\mathcal{H}_1\otimes ...\otimes\mathcal{H}_{N+1}$. For a given pure state we trace out the degrees of freedom in $\mathcal{H}_{N+1}$ to obtain an $N-$partite mixed state $\rho_N$. Then we compute the entropy of each of the $N$ individual subsystems of $\rho_N$, all the pairs, triplets and so on, up to the entropy of the union of all parties (the entropy of $\rho_N$) and we arrange these numbers into a vector in ``entropy space'' $\mathbb{R}^{2^N-1}$. Consider now the set of all such vectors obtained from all possible initial pure states in all possible Hilbert spaces with the previously mentioned tensor product structure. The subset of etntropy space so obtained has the structure of a convex cone and is known as the \textit{quantum entropy cone} ($\mathcal{QC}_N$).

For $N=2,3$ the cones are known to be polyhedral and they are then specified by a finite number of linear inequalities. Any polyhedral cone can equivalently be described by the list of its \textit{extremal rays} since any vector inside the cone or on its boundary can be obtained from a conical combination of them.\footnote{A conical combination is a linear combination with non-negative coefficients.}

To clarify the construction, and for the purpose of the later discussion, let us review the examples of the cones $\mathcal{QC}_2$ and $\mathcal{QC}_3$. In the $N=2$ case the inequalities that determine the quantum cone are simply subadditivity ($\mathcal{SA}$) and Araki-Lieb ($\mathcal{AL}$)
\begin{align}
\mathcal{SA}:&\qquad S_A+S_B\geq S_{AB} \,,\nonumber\\
\mathcal{AL}:&\qquad S_A+S_{AB}\geq S_B \,.
\end{align}
Note that even if the two inequalities are physically equivalent, in the sense that one implies the other, both are necessary for the construction of the cone because they correspond to different facets. For the same reason one should also include both versions of $\mathcal{AL}$ obtained by swapping $A$ and $B$. The extremal rays are
\begin{equation}
(S_A,S_B,S_{AB})\in\{(1,0,1),(0,1,1),(1,1,0)\}\,,
\end{equation}
corresponding to obvious quantum states. These rays are actually equivalent since they are mapped to each other by permutations of the subsystems. Here we list all of them just for clarity, in the following we will only focus on inequivalent rays.

To obtain the $N=3$ cone we can imagine that we first build a ``candidate cone'' obtained from all possible versions of the $N=2$ inequalities for three subsystems $A,B,C$. We then cut away parts of this cone by slicing along the hyperplanes corresponding to all new, genuinely $3-$partite, inequalities.\footnote{For example for $\mathcal{SA}$ one also has $S_A+S_C\geq S_{AC}$, $S_A+S_{BC}\geq S_{ABC}$ and various permutations. However this description is redundant and some of the inequalities can be removed.} These are strong subadditivity ($\mathcal{SSA}$) and weak monotonicity ($\mathcal{WM}$)\footnote{Again the two inequalities are equivalent but both should be considered.}
\begin{align}
\mathcal{SSA}:&\qquad S_{AB}+S_{BC}\geq S_{B}+S_{ABC} \,, \nonumber\\
\mathcal{WM}:&\qquad S_{AB}+S_{BC}\geq S_A+S_C \,.
\end{align}
The extremal rays are then, up to permutations
\begin{align}
&(S_A,S_B,S_C,S_{AB},S_{AC},S_{BC},S_{ABC})\in\nonumber\\
&\{(1,1,0,0,1,1,0),(1,1,1,2,2,2,1),(1,1,1,1,1,1,1)\} \,.
\label{eq:rays3}
\end{align}
The first ray simply corresponds to a Bell pair for $AB$, the second is a four-qutrit stabilizer state and the last one is a GHZ state of four qubits.\footnote{The expressions of the states realizing the second and third extremal rays are $\sum_{i,j=0}^2\ket{i,j,i+j,i+2j}$ and $\ket{0000}+\ket{1111}$ respectively (up to a normalization factor).}

For $N\geq 4$ new inequalities are known, but the full structure of the cones is not known. However, they are conjectured to be non-polyhedral \cite{Matus:2007aa}.

\subsection{The Metric Entropy Cone ($\mathcal{MC}_N$)}
\label{subsec:areacone}

Having introduced the general concept of the quantum entropy cone we can now define, following \citep{Bao}, an analogous object for areas of minimal surfaces in a geometric set-up inspired by holography. Consider an arbitrary $d+1$ dimensional manifold with $(N+1)$ boundaries $\mathcal{M}_{N+1}$, with a metric which is asymptotically AdS$_{d+1}$ approaching the boundaries. Since we are restricting to the RT prescription, rather than the more general HRT \citep{HRT}, we will also assume that the metric has a time reflection symmetry at $t=0$ and consider the $d$ dimensional manifold $\widetilde{\mathcal{M}}_{N+1}$ corresponding to the $t=0$ slice of $\mathcal{M}_{N+1}$. We are interested in the area of minimal surfaces homologous to regions specified on the boundaries of $\widetilde{\mathcal{M}}_{N+1}$. In this set-up it is natural to think that such a geometry is dual to a pure state of a tensor product of $N+1$ CFTs living on the boundaries, but for the moment we do not make this assumption. The \textit{metric entropy cone}($\mathcal{MC}_N$) is then defined as the region in the space of ``area vectors'' $\mathbb{R}^{2^N-1}$ spanned by varying the topology and the metric of $\mathcal{M}_{d+1}$, as well as the choice of the $N$ subregions. This is precisely the definition of \citep{Bao} for the ``holographic entropy cone''.

For the case $N=2$, it follows from the ``RT proof'' of $\mathcal{SA}$, and from the fact that one can construct a geometry that realizes the extremal rays (a two boundary wormhole), that $\mathcal{QC}_2\equiv\mathcal{MC}_2$. We will focus on the more interesting cases $N=3,4$. To construct the metric entropy cone for $N=3$ one should include monogamy of mutual information ($\mathcal{MMI}$) to the list of inequalities\footnote{Interestingly $\mathcal{SSA}$ can be removed from the list. The reason is that it is redundant as it can be obtained from $\mathcal{MMI}$ and $\mathcal{SA}$.}
\begin{equation}
\mathcal{MMI}:\qquad S_{AB}+S_{AC}+S_{BC}\geq S_A+S_B+S_C+S_{ABC} \,.
\end{equation}
The list of extremal rays is then updated to
\begin{align}
(S_A,S_B,S_C,S_{AB},S_{AC},S_{BC},S_{ABC})\in\{(1,1,0,0,1,1,0),(1,1,1,2,2,2,1)\} \,.
\label{eq:rays3area}
\end{align}
Comparing \eqref{eq:rays3area} with \eqref{eq:rays3} one sees that the net effect of including $\mathcal{MMI}$ simply is the removal of the ray corresponding to the GHZ state. Geometrically, the first extremal ray in \eqref{eq:rays3} is realized by the disjoint union of a two boundary wormhole and two copies of empty AdS. The second ray is instead a $4$-boundary wormhole which we will study in \S\ref{subsec:three}.

In the $N=4$ case it has already been shown in \citep{Bao} that there are no new inequalities implied by the RT formula. To see that this is the case one starts again with all the previous inequalities for fewer parties and consider all possible versions for the $4-$partite case. From this list of inequalities one can then construct a cone, which is a ``candidate'' for $\mathcal{MC}_4$, and can extract its extremal rays. The result is
\begin{align}
&(S_A,S_B,S_C,S_D,S_{AB},S_{AC},S_{AD},S_{BC},S_{BD},S_{CD},S_{ABC},S_{ABD},S_{ACD},S_{BCD},S_{ABCD})\nonumber\\
&\in\{(0,0,0,1,0,0,1,0,1,1,0,1,1,1,1), (0,1,1,1,1,1,1,2, 2, 2, 2, 2, 2,1,1),\nonumber\\
&(1,1,1,1,2,2,2,2,2,2,3,3,3,3,2)\}\,.
\label{eq:rays4}
\end{align}
The first two rays are again inheritated from those of the previous cones. The last one instead is again a multiboundary geometry which we will investigate in \S\ref{subsec:four}. Since any extremal ray of this candidate cone can in fact be realized by some geometry, it follows from convexity that any other ray inside the cone can also be realized geometrically. This proves that for four regions there cannot be new RT inequalities and that the candidate previously constructed is in fact $\mathcal{MC}_4$. For $N>4$ the metric entropy cones are not known, but it was proved in \citep{Bao} that they are all polyhedral. Furthermore, it was proved in \cite{Hayden:2016aa} that $\mathcal{MC}_N\subseteq\mathcal{QC}_N$ for all $N$, also justifying the usage of the word ``entropy''.

\subsection{The Ryu-Takayanagi Cone ($\mathcal{RTC}_N$)}
\label{subsec:holographiccone}

The construction presented in the previous section was completely geometric and although it was motivated by holography it did not really require it. However, since we are interested in the set of constraints imposed by the RT formula on the space of CFT states with classical bulk duals, we want to be able to interpret the ``areas'' which appeared in the metric entropy cone as von Neumann entropies of regions in field theory.\footnote{In these work we only consider the leading contribution to the entropy in the large N limit.} For this to be true one needs to further constraint the allowed geometries for which one computes areas and impose that such geometries are in fact dual to some CFT state. We then define the \textit{Ryu-Takayanagi cone ($\mathcal{RTC}_N$)} as the cone spanned by all holographic states, for an arbitrary number of CFTs and all possible choices of the $N$ regions. This is a convex cone, since given any two rays it contains any conical combination of them, obtained by rescaling the metric and taking the tensor product of the corresponding two states.\footnote{As for the metric entropy cone, we restrict attention to geometries with a moment of time symmetry, although one could similarly define a more general HRT cone.}

From the previous definitions, and the result of \cite{Hayden:2016aa}, it follows that for any $N$
\begin{align}
\mathcal{RTC}_N\subseteq\mathcal{MC}_N\subseteq\mathcal{QC}_N\,.
\label{eq:inclusions}
\end{align}
The case $N=2$ is trivial since all the cones coincide, while for $N\geq 3$ the second inclusion in \eqref{eq:inclusions} is strict as a consequence of $\mathcal{MMI}$. On the other hand the question of whether the first inclusion in \eqref{eq:inclusions} is strict is the focus of the next section (for $N=3,4$). 


\section{Constructing holographic geometries for the extremal rays}
\label{sec:rays}

In this section we explore the relation between the previously defined metric entropy and \textit{RT} cones in the particular cases of $N=3$ and $N=4$. To prove that the RT cone coincides with the metric entropy cone, one needs to find holographic CFT states dual to the geometries that realize the extremal rays of the metric entropy cone. Since the metric entropy cone was defined for the RT formula (instead of HRT) we restrict to spacetimes with a time reflection symmetry. We focus on three dimensional gravity such that all possible solutions of Einstein's equations are locally AdS$_3$ and can be obtained by its quotients. The $t=0$ slice of the full spacetime will then be a Riemann surface $\Sigma_{N+1}$ with $N+1$ boundaries. Finally, following \citep{Bao}, we assume that the $N$ boundary regions for which we compute areas of RT surfaces are entire boundaries.

The solutions we are interested in are then multiboundary wormholes with $N+1$ asymptotic boundaries with a set of constraints on the size of the horizons and all internal cycles such that the geometries correspond to extremal rays of the metric entropy cone. We seek to find states in the tensor product of $N+1$ CFTs which are dual to such wormholes and can be obtained from a bulk Euclidean path integral via the Hartle-Hawking construction. For both the $N=3$ and $N=4$ cases we fail to find CFT states that meet all the requirements.

\subsection{Handlebody solutions}
\label{subsec:handlebodies}

A particularly interesting class of Euclidean solutions of three dimensional Einstein gravity with negative cosmological constant are the so-called handlebody solutions, which can be thought as compact Riemann surfaces ``filled in'' with hyperbolic space. While these are not the only solutions for a particular set of boundary conditions, it has been conjectured that the non-handlebody solutions are always sub-dominant \cite{Yin}. Our numerics confirm this explicitly in certain contexts (see \S\ref{sec:reduction} for more details).

As first proposed in \cite{Krasnov1, Krasnov2} (see also \citep{Skenderis:2009aa}), one can interpret such a solution for $t_E\in(-\infty,0)$ as a saddle point of the bulk Euclidean path integral. However, for a given compact Riemann surface, different handlebodies can be obtained by different fillings and correspond to different phases \cite{MBW1}. If this saddle is the one with least action then by the Hartle-Hawking construction it provides the bulk state at $t=0$ and via holography an approximation to the dual CFT state computed by the field theory path integral. For our purposes we must then check that the correct phase dominates in the region of moduli space where the entanglement entropies satisfy the relations determined by the extremal ray. Recently developed techniques \citep{MRW} allow us to study these solutions numerically even in the case of three or more boundaries. While evaluating the action for these geometries is involved, numerical evidence from our work and \citep{MRW} suggests that a useful heuristic is that filling in the bulk along smaller boundary cycles costs less action than filling along larger boundary cycles. A related effect is that geometries with small internal bulk cycles tend to be subdominant in the path integral. This provides a coarse way to understand which phase dominates at a given point in moduli space.

An arbitrary handlebody solution can be obtained as a quotient of three dimensional hyperbolic space $\mathbb{H}^3$ (i.e. Euclidean AdS$_3$). A particular quotient can be specified by its action on the boundary Riemann surface, with the action extended into the bulk by geodesics. Explicitly, to construct a quotient of $\mathbb{H}^3$ with a genus $g$ Riemann surface its the boundary, one chooses $g$ pairs of non-intersecting circles $C_k,C_k'$ which divide the Riemann sphere into a region ``inside'' and ``outside.'' The set of M\"obius transformations $L_k$ mapping $C_k \mapsto C_k'$ are then used to define the particular quotient of $\mathbb{H}^3$, with the action in the bulk defined by an extension along geodesics. The Riemann sphere projected onto $\mathbb R^2$ with the specifications of a given set of circles is referred to as a Schottky domain $D$ (see \cite{Krasnov1, Skenderis:2009aa} for more details).

In practice, it is difficult to determine the handlebody corresponding to a particular quotient of the Riemann sphere. Instead it is more convenient to construct a Schottky domain for a particular Riemann surface by choosing $g$ cycles on the boundary to be made contractible in the bulk.\footnote{One way to classify the possibilities is to first construct a basis of cycles $\{\alpha_i, \beta_i\}$ such that $\alpha_i \cap \beta_j = \delta_{ij}$. Letting the cycles $\{\alpha_i\}$ be contractible in the bulk defines a particular handlebody solution. One can then act on this basis with an element of the mapping class group and choose the resulting $\alpha$ cycles to be contractible. Acting with all elements of the mapping class group generates a set of handlebody phases for a given boundary surface.} Cutting the Riemann surface along these $g$ cycles gives a Riemann sphere with $g$ sets of circles identified, which can then be projected into the plane to define the Schottky domain. Determining the moduli of the Riemann surface corresponding to a particular domain is done numerically for $g>1$, and we must solve this moduli matching problem to compare solutions with given boundary conditions.

In order to compare the actions for different solutions, we must first choose a conformal frame on the boundary. The standard choice is $R_\text{bndy}=-2$. As shown in \cite{Krasnov1}, if we write the boundary metric as
\ban{
ds^2 = e^{2\phi(w,\bar w)}dw d\bar w\,.
}
then this choice of conformal frame corresponds to the solution of the Liouville equation with an additional non-trivial condition
\ban{
\nabla^2 \phi = e^{2\phi}\quad\,, \text{ with } \quad
\phi(L_i(w)) = \phi(w) -\frac 12 \log \left| L'_i(w)\right|^2 \,.
}
We will use the numerical methods described in \cite{MRW} to solve this differential equation for a given domain.

Additionally, as shown in \citep{Krasnov1}, the evaluation of the Einstein-Hilbert action for a particular solution can be written in terms of the Takhtajan-Zograf action \citep{TZ} for the scalar field $\phi$. As explained in \cite{MRW}, if we define $R_k$ to be the radius of $C_k$ and $\Delta_k$ as the distance between the center of $C_k$ and the point $w_\infty^{(k)}$ mapped to $\infty$ by $L_k$, this action reduces to
\ban{
I_{TZ}[\phi] = \int_D d^2 w\left( \left(\nabla \phi\right)^2 + e^{2\phi} \right) + \sum_k \left(\int_{C_k} 4 \phi\, d\theta_\infty^{(k)}  - 4 \pi \log \left |R_k^2 - \Delta_k^2 \right|\right)\,.
}
where $\theta_\infty^{(k)}$ is the angle measured from the point $w_{\infty}^{(k)}$. More details on the numerical evaluation of this action can be found in \S\ref{sec:reduction}.

Finally, we will have to impose on a handlebody solution a set of constraints which guarantee that the entanglement entropies of entire boundaries (and their unions) match the values of a particular extremal ray of the metric entropy cone. Since the RT formula relates such entropies to geodesic lengths, we need to relate these geodesics lengths to the quotient which defines a particular handlebody. This can be done using the results of \cite{Maxfield3D}, which showed that the length of a geodesic in the homology class corresponding to the action of a M\"obius transformation $L$ is given by
\ban{
\ell (L) = \cosh^{-1} \left[ \frac {\Tr L }2 \right] \,.
}
In the next sections we will use all of this technology to evaluate the on-shell action for the solutions corresponding to the desired extremal rays of the $N=3,4$ metric entropy cones.

\subsection{Four party extremal rays}
\label{subsec:four}

%
\begin{figure}[tb]
\centering
{\includegraphics[trim=0 6.2cm 0 4cm,clip=true,width=0.75\textwidth]{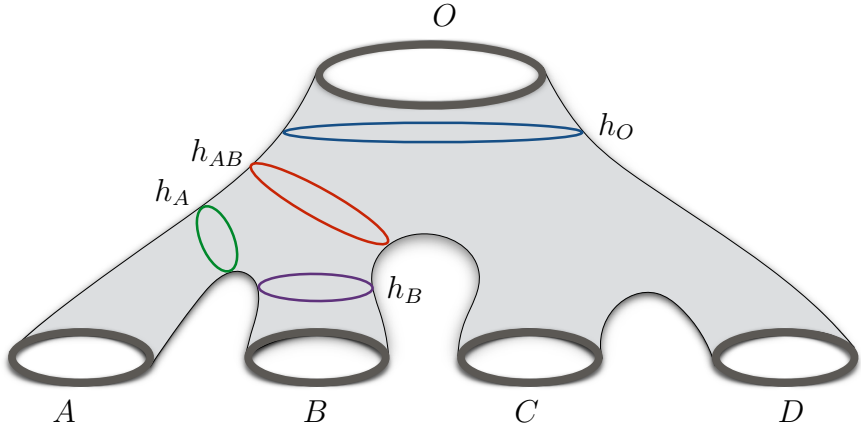}}
\put(-305,-6){\makebox(0,0){$A$}}
\put(-210,-6){\makebox(0,0){$B$}}
\put(-130,-6){\makebox(0,0){$C$}}
\put(-30,-6){\makebox(0,0){$D$}}
\put(-161,144){\makebox(0,0){$O$}}
\put(-264,77){\makebox(0,0){$h_A$}}
\put(-176,40){\makebox(0,0){$h_B$}}
\put(-247,92){\makebox(0,0){$h_{AB}$}}
\put(-96,102){\makebox(0,0){$h_O$}}
\caption{The geometry (at $t=0$) corresponding to the four party extremal ray of interest. The sizes of the horizons and the internal cycles are fixed in order to obtain the correct entropies, see the main text.}
\label{fig:four}
\end{figure}
\begin{figure}[b!]
\centering
\begin{subfigure}{0.49\textwidth}
\includegraphics[trim=9cm 7cm 9cm 7cm,clip=true,width=\textwidth]{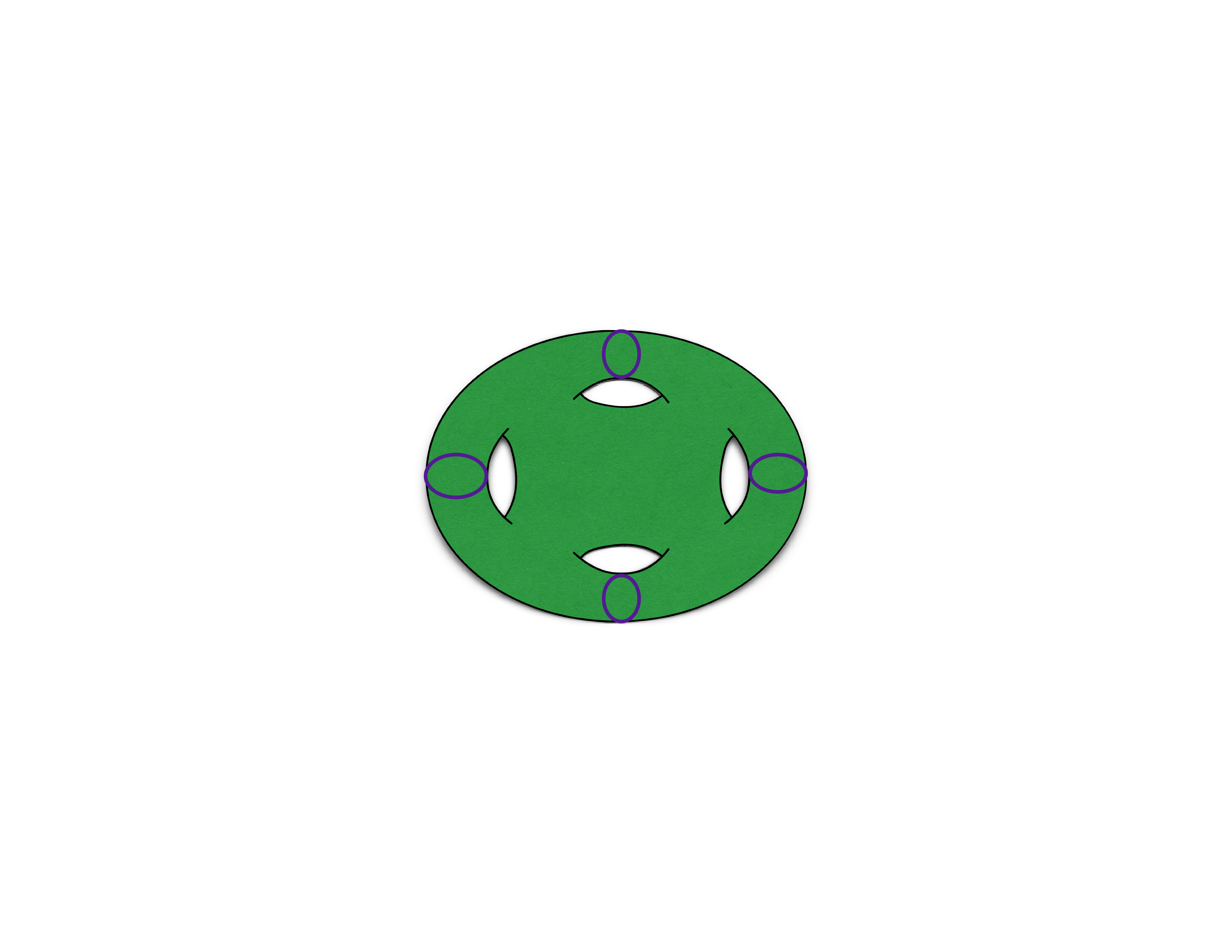}
\put(-51,84){\makebox(0,0){\footnotesize{$A$}}}
\put(-106,124){\makebox(0,0){\footnotesize{$B$}}}
\put(-164,84){\makebox(0,0){\footnotesize{$C$}}}
\put(-106,42){\makebox(0,0){\footnotesize{$D$}}}
\put(-50,148){\makebox(0,0){\footnotesize{$O$}}}
\subcaption{The connected phase.}
\label{fig:conn4}
\end{subfigure}
\hfill
\begin{subfigure}{0.49\textwidth}
\includegraphics[trim=2.5cm 3cm 2.5cm 3cm,clip=true,width=\textwidth]{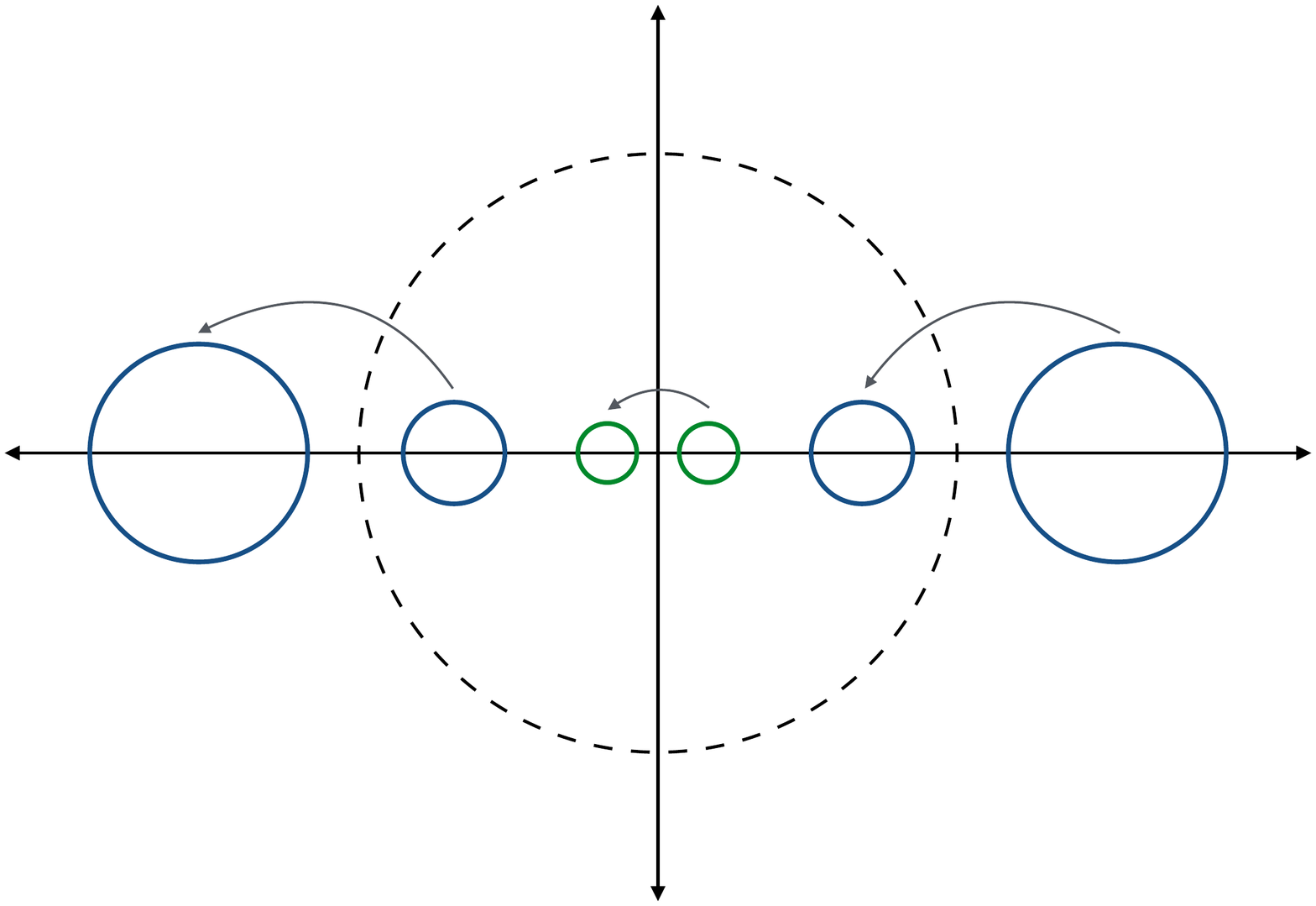}
\put(-50,105){\makebox(0,0){\footnotesize{$L_1$}}}
\put(-100,89){\makebox(0,0){\footnotesize{$L_2$}}}
\put(-162,105){\makebox(0,0){\footnotesize{$L_3$}}}
\subcaption{The associated Schottky domain, with two further circles not drawn, but implied by requiring inversion symmetry across the dashed circle}.
\label{fig:schottky_conn4}
\end{subfigure}

\begin{subfigure}{0.49\textwidth}
\includegraphics[trim=9.3cm 7cm 8.7cm 7cm,clip=true,width=\textwidth]{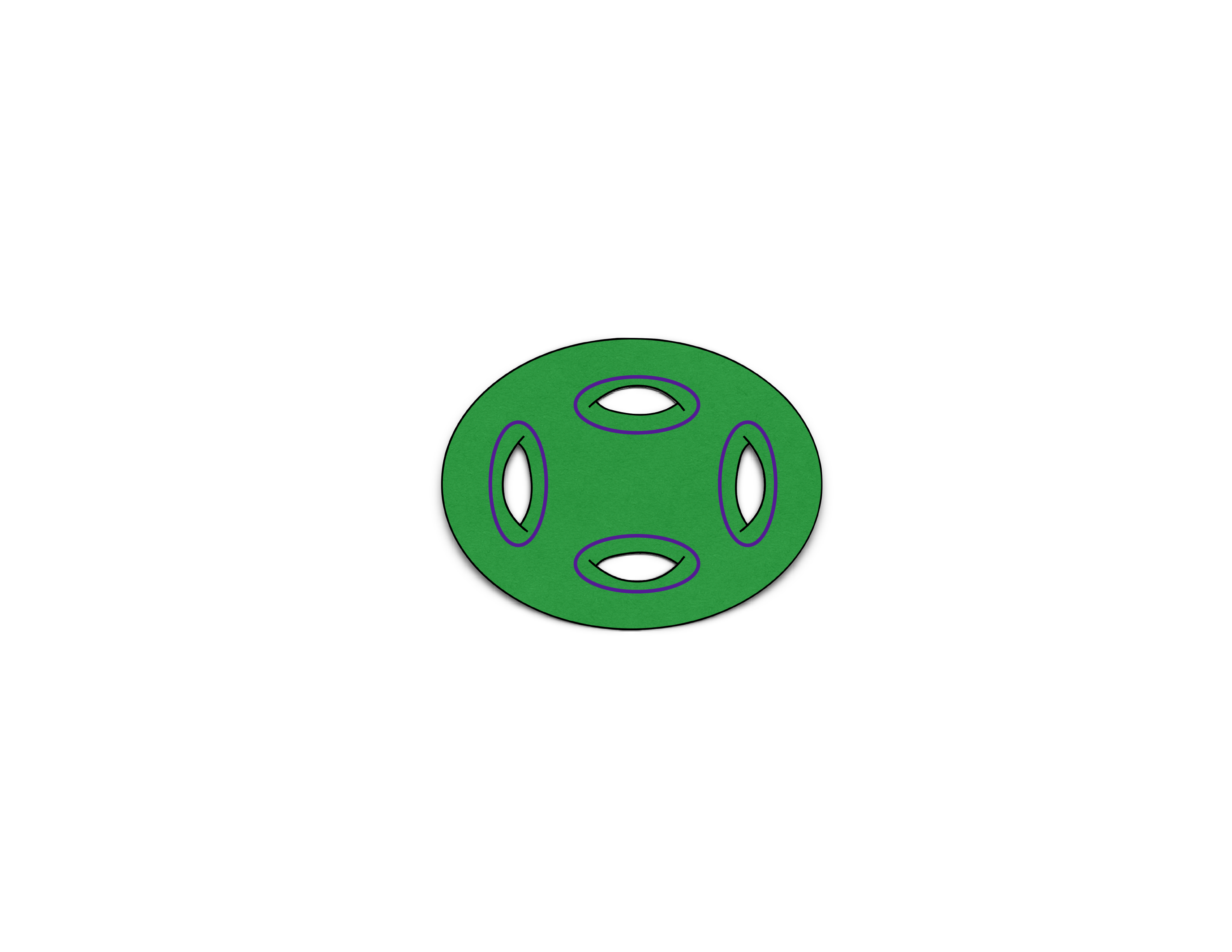}
\put(-50,82){\makebox(0,0){\footnotesize{$A$}}}
\put(-105,120){\makebox(0,0){\footnotesize{$B$}}}
\put(-164,82){\makebox(0,0){\footnotesize{$C$}}}
\put(-105,38){\makebox(0,0){\footnotesize{$D$}}}
\put(-52,145){\makebox(0,0){\footnotesize{$O$}}}
\subcaption{The disconnected phase.}
\label{fig:disconn4}
\end{subfigure}
\hfill
\begin{subfigure}{0.49\textwidth}
\includegraphics[trim=2cm 1cm 2cm 1cm,clip=true,width=\textwidth]{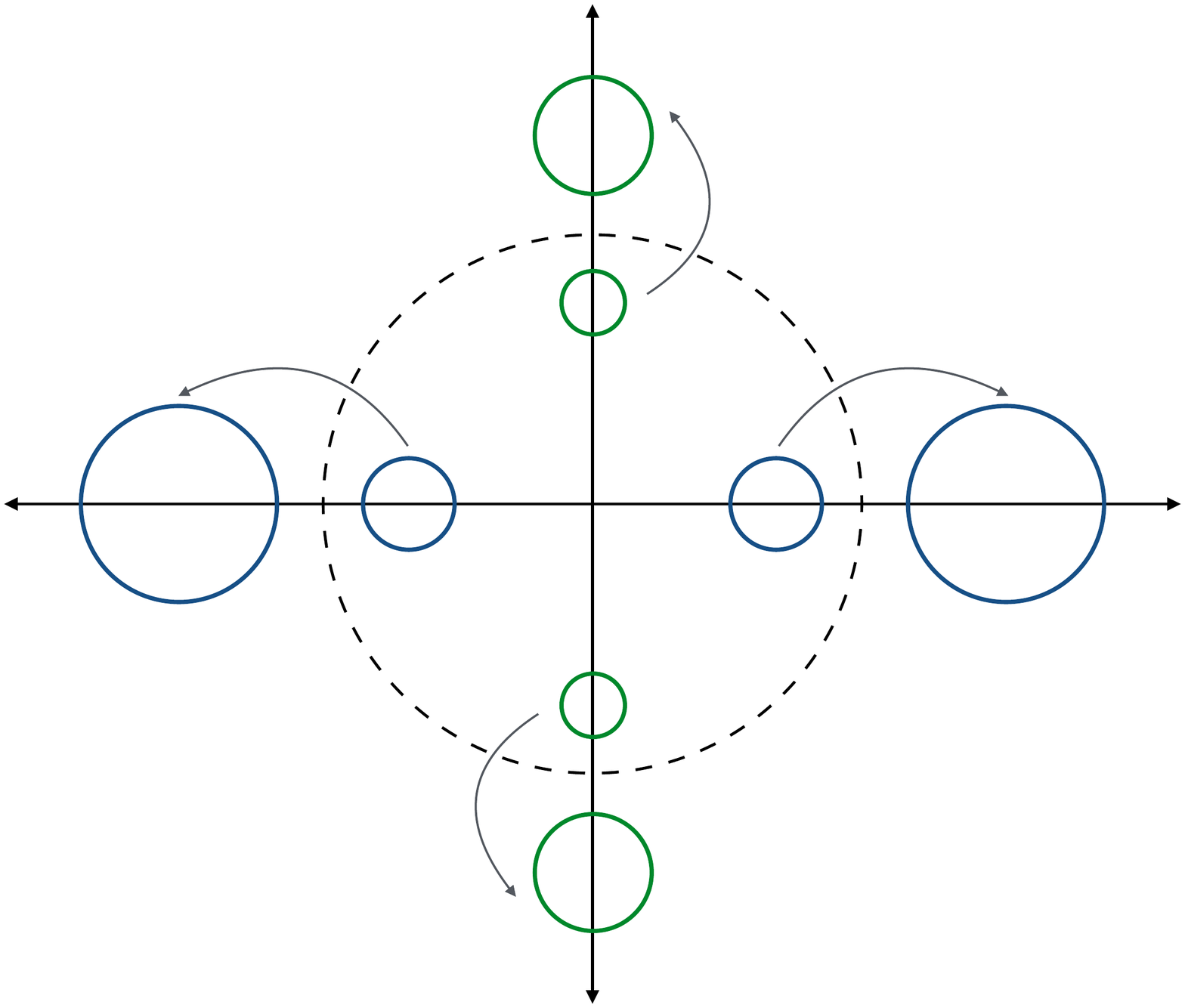}
\put(-50,120){\makebox(0,0){\footnotesize{$L_1$}}}
\put(-80,140){\makebox(0,0){\footnotesize{$L_2$}}}
\put(-162,120){\makebox(0,0){\footnotesize{$L_3$}}}
\put(-135,30){\makebox(0,0){\footnotesize{$L_4$}}}
\subcaption{The associated Schottky domain.}
\label{fig:schottky_disconn4}
\end{subfigure}
\caption{The choices of contractible cycles corresponding to the connected and disconnected phases and the corresponding Schottky domains.}
\label{fig:geom}
\end{figure}

The extremal rays of $\mathcal{MC}_4$ where listed in \eqref{eq:4ray}. The first two rays are inherited from the three party cone $\mathcal{MC}_3$ and we will ignore them, we will briefly comment on the three party case in the next section. Here instead we focus on the only new extremal ray
\begin{align}
&(S_A,S_B,S_C,S_D,S_{AB},S_{AC},S_{AD},S_{BC},S_{BD},S_{CD},S_{ABC},S_{ABD},S_{ACD},S_{BCD},S_{ABCD})\nonumber\\
&=(1,1,1,1,2,2,2,2,2,2,3,3,3,3,2)\,.
\label{eq:4ray}
\end{align}
As already shown in \citep{Bao}, this ray can be realized by the multiboundary wormhole geometry drawn in Fig.~\ref{fig:four}, with the following conditions on the horizon lengths
\begin{align}
l=|h_A| = |h_B|=|h_C|=|h_D| = \frac{1}{2}|h_O|\,.
\label{eq:constraints1}
\end{align}
and with the additional constraint that for any internal cycle $\gamma$ homologous to a union of $n$ of the four boundaries one has
\begin{align}
|\gamma|\geq nl\,.
\label{eq:constraints2}
\end{align}
For example we should impose $|h_{AB}| \geq 2l$.

\begin{figure}[b!]
\centering
\begin{subfigure}{0.49\textwidth}
\includegraphics[width=\textwidth]{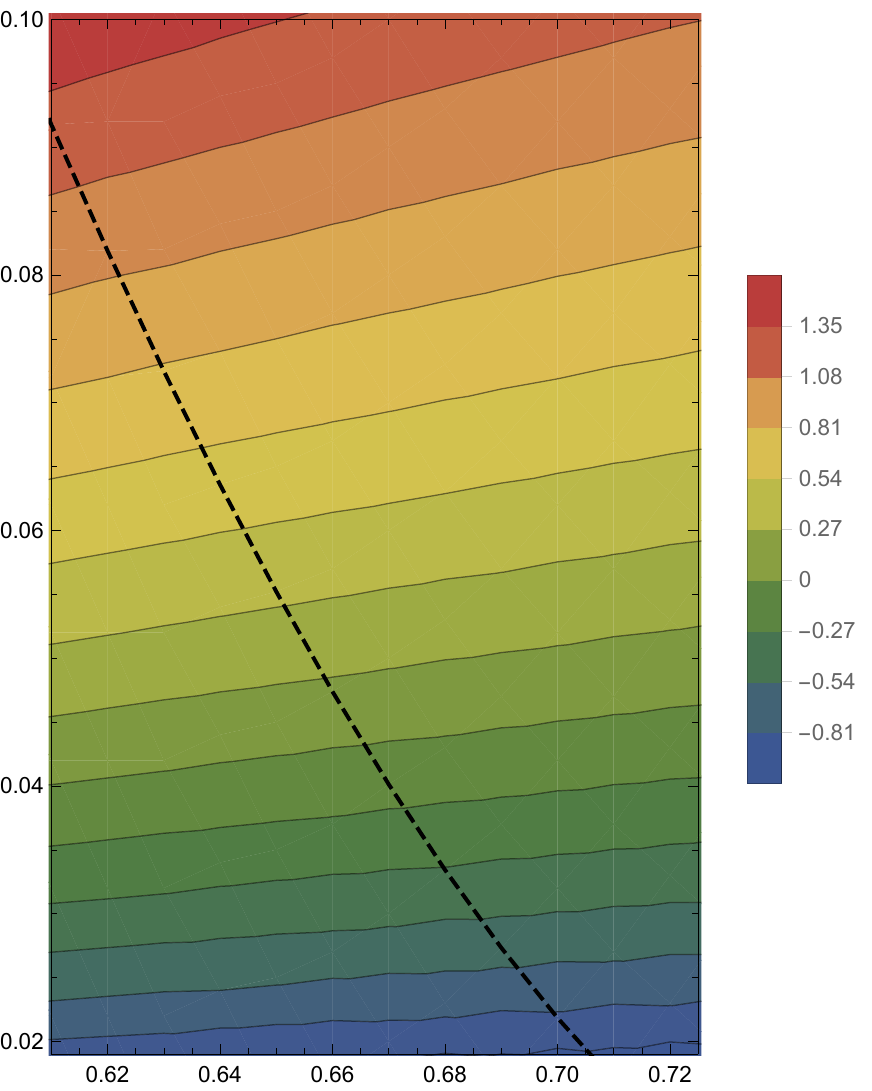}
\put(-212,257){\makebox(0,0){\footnotesize{$\rho$}}}
\put(-36,4){\makebox(0,0){\footnotesize{$\phi$}}}
\subcaption{$I_{\text{con}}-I_{\text{dis}}$}
\label{fig:result4a}
\end{subfigure}
\hfill
\begin{subfigure}{0.49\textwidth}
\includegraphics[width=\textwidth]{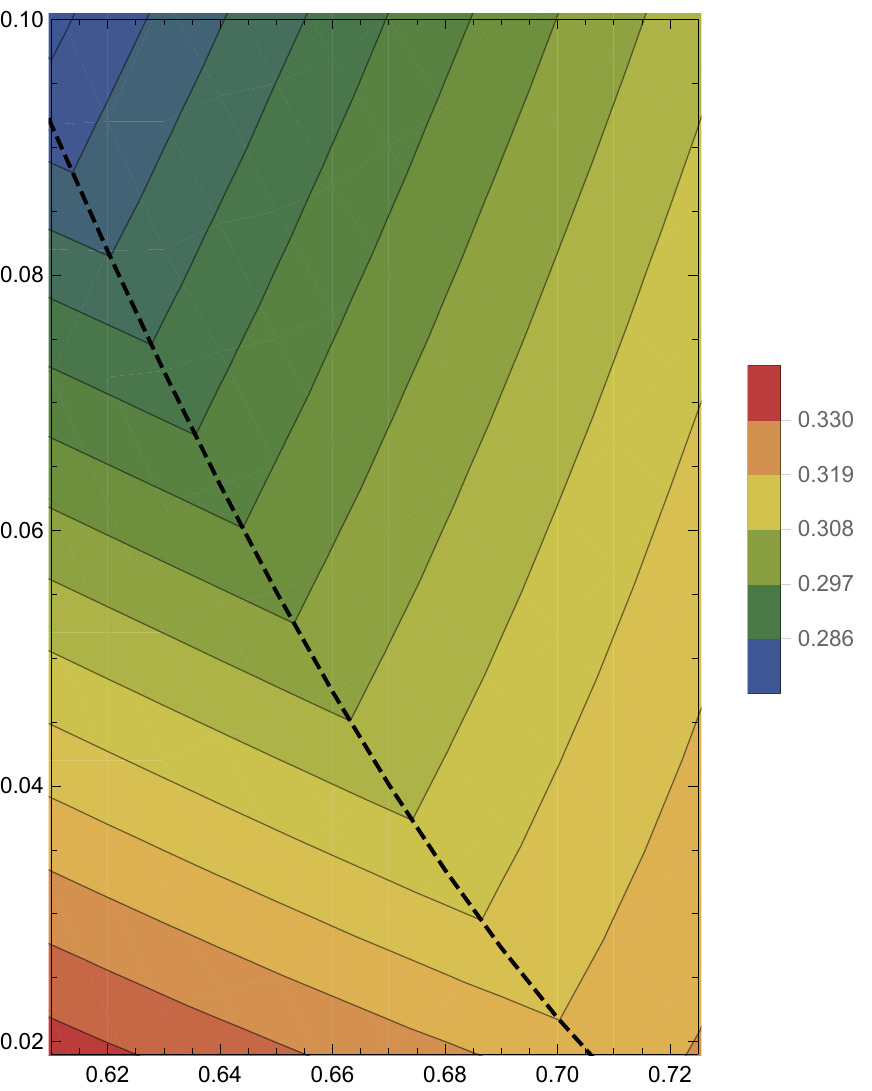}
\put(-212,257){\makebox(0,0){\footnotesize{$\rho$}}}
\put(-36,4){\makebox(0,0){\footnotesize{$\phi$}}}
\subcaption{$\max_\gamma(1-\frac{|\gamma|}{n\,l})$}
\label{fig:result4b}
\end{subfigure}
\caption{Comparison of the actions for the connected and disconnected phases in the symmetric configuration. The parameters ($\phi,\rho$) are the angular position and size of $C_1$ on the Riemann sphere. The dashed line corresponds to the particular configuration with an additional $\pi/2$ rotational symmetry about the axis orthogonal to the plane of the page in Fig.~\ref{fig:conn4} and Fig.~\ref{fig:disconn4} (see also Fig.~\ref{fig:resultSCDequal}).}
\label{fig:plots}
\end{figure}

We wish to evaluate the Euclidean action for this solution, which we refer to as the \textit{connected phase}, and compare it to the action evaluated for a solution where the surface of time-reflection symmetry in the bulk does not connect any pair of boundaries (which we will simply call the \textit{disconnected phase}).\footnote{While in principle there are multiple phases where the surface of time reflection symmetry in the bulk is disconnected, for our analysis it is sufficient to focus only on this particular phase.} If the connected phase is the one that minimizes the action then as explained earlier we can conclude that  in the limit of small bulk Newton constant $G$ (and thus large CFT central charge $c$) it is in fact dual to a field theory state. As anticipated, we will find evidence that this is not the case. The constraints \eqref{eq:constraints2} require that the cycles homologous to  single boundaries are small compared to the other internal cycles. Intuitively then, the action favors the disconnected phase, where the wormhole ``pinches off'' and these cycles become contractible in the bulk.

The wormhole geometry of Fig.~\ref{fig:four} can be obtained starting from a closed Riemann surface of genus $g=4$. Different handlebody solutions with the same boundary  correspond to different choices for the cycles which are contractible in the bulk. The choice that gives the connected phase is shown in Fig.~\ref{fig:conn4} and the corresponding Schottky domain in Fig.~\ref{fig:schottky_conn4}. In the case of the disconnected phase (see Fig.~\ref{fig:disconn4} and Fig.~\ref{fig:schottky_disconn4}), the bulk time slice consists of five disconnected copies of $\mathbb{H}^2$; the mutual information between boundaries vanishes and the constraints \eqref{eq:constraints1} are not satisfied.

We want to compare the action for these two phases in various regions of the moduli space for the associated genus-4 Riemann surface.  We focus on a subspace defined by imposing three $\mathbb Z_2$ symmetries corresponding (for Fig.~\ref{fig:conn4} and Fig.~\ref{fig:disconn4}) to reflections in the plane of the page and across both the horizontal and vertical axes. In both phases considered, the contractible cycles are homotopic to curves that are fixed by a subset of these symmetries. Combined with the constraints \eqref{eq:constraints1}, these conditions reduce the moduli space to two real dimensions. In Fig.~\ref{fig:result4a} we plot the difference between disconnected ($I_{\text{dis}}$) and connected ($I_{\text{con}}$) actions and in Fig.~\ref{fig:result4b} we show $\max_\gamma(1-\frac{|\gamma|}{n\,l})$; i.e. the maximal violation of the constraints \eqref{eq:constraints2}. As announced before, one can clearly see that the constraints are violated when the connected phase has smaller action (bluer regions of Fig.~\ref{fig:result4b}).  Indeed, the region displayed in Fig.~\ref{fig:result4b} contains no points where the constraints are all satisfied.

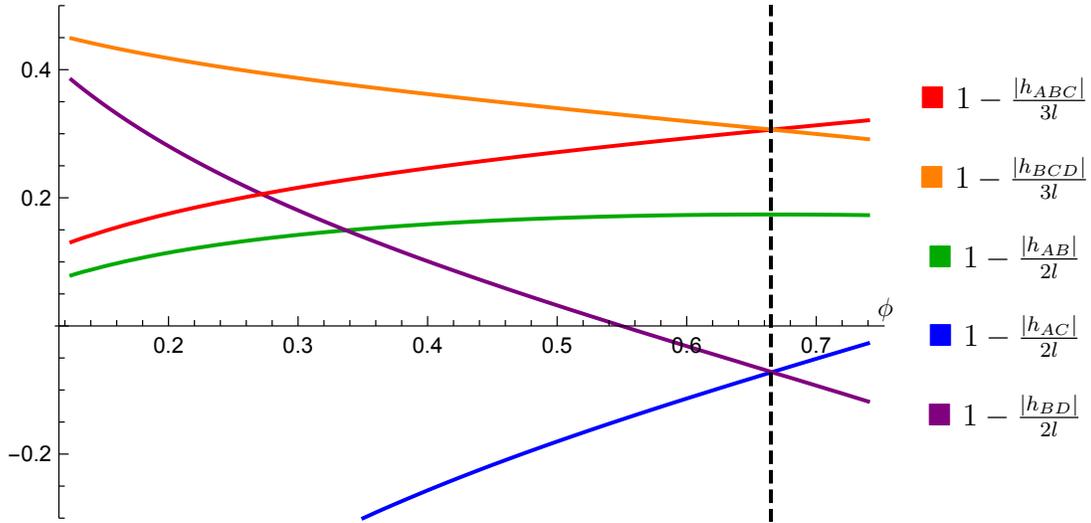
\begin{figure}[t!]
{\includegraphics[width=0.75\textwidth]{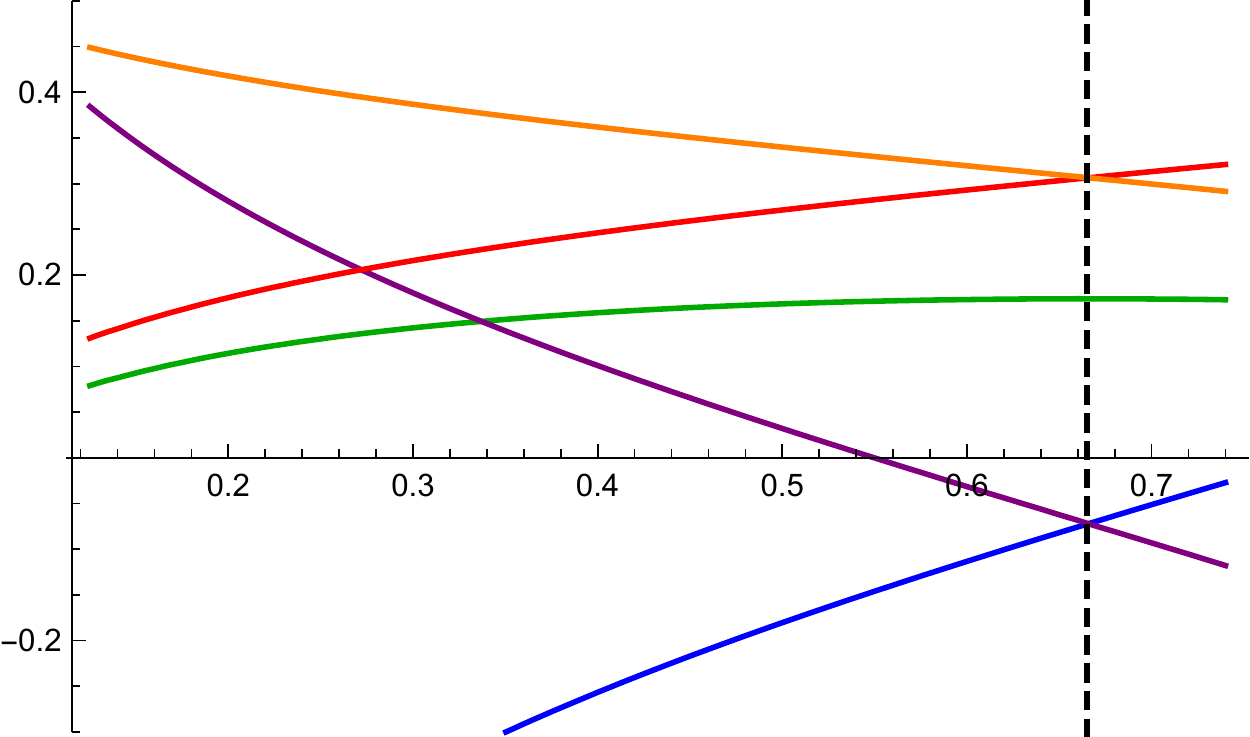}}
\centering
\hspace*{2.5cm}
\put(-30,160){\makebox(0,0){$\textcolor[RGB]{255,0,0}{\blacksquare}\,\,1-\frac{|h_{ABC}|}{3l}$}}
\put(-30,130){\makebox(0,0){$\textcolor[RGB]{255,128,0}{\blacksquare}\,\,1-\frac{|h_{BCD}|}{3l}$}}
\put(-30,100){\makebox(0,0){$\textcolor[RGB]{0,170,0}{\blacksquare}\,\,1-\frac{|h_{AB}|}{2l}$}}
\put(-30,70){\makebox(0,0){$\textcolor[RGB]{0,0,255}{\blacksquare}\,\,1-\frac{|h_{AC}|}{2l}$}}
\put(-30,40){\makebox(0,0){$\textcolor[RGB]{128,0,128}{\blacksquare}\,\,1-\frac{|h_{BD}|}{2l}$}}
\put(-75,80){\makebox(0,0){\footnotesize{$\phi$}}}
\caption{The constraints \eqref{eq:constraints2} in the subspace where the connected and disconnected actions are equal.  The dashed line indicates the point with the additional $\pi/2$ rotational symmetry discussed above.}
\label{fig:resultSCDequal}
\end{figure}

Now the particular region of moduli space studied in Fig.~\ref{fig:plots} was chosen in an ad hoc way for numerical convenience.  To perform a more targeted analysis, we restrict attention to the 1-dimensional subspace of moduli space where the connected and disconnected actions are equal.  We can then search along this curve for a point where the constraints \eqref{eq:constraints2} are satisfied. Though we have no rigorous proof, it is natural to presume that it is easier to satisfy the constraints along this line than in the region where the connected phase dominates strongly. The plot shown in Fig.~\ref{fig:resultSCDequal} suggests that no such point exists. Toward the left, the $h_{BD}$ and $h_{BCD}$ cycles become too small, while the $h_{ABC}$, $h_{AB}$, and $h_{AC}$ cycles become too small toward the right.

\begin{figure}[b!]
\centering
\begin{subfigure}{0.49\textwidth}
\includegraphics[trim=3.5cm 13cm 3.5cm 4.5cm,clip=true,width=\textwidth]{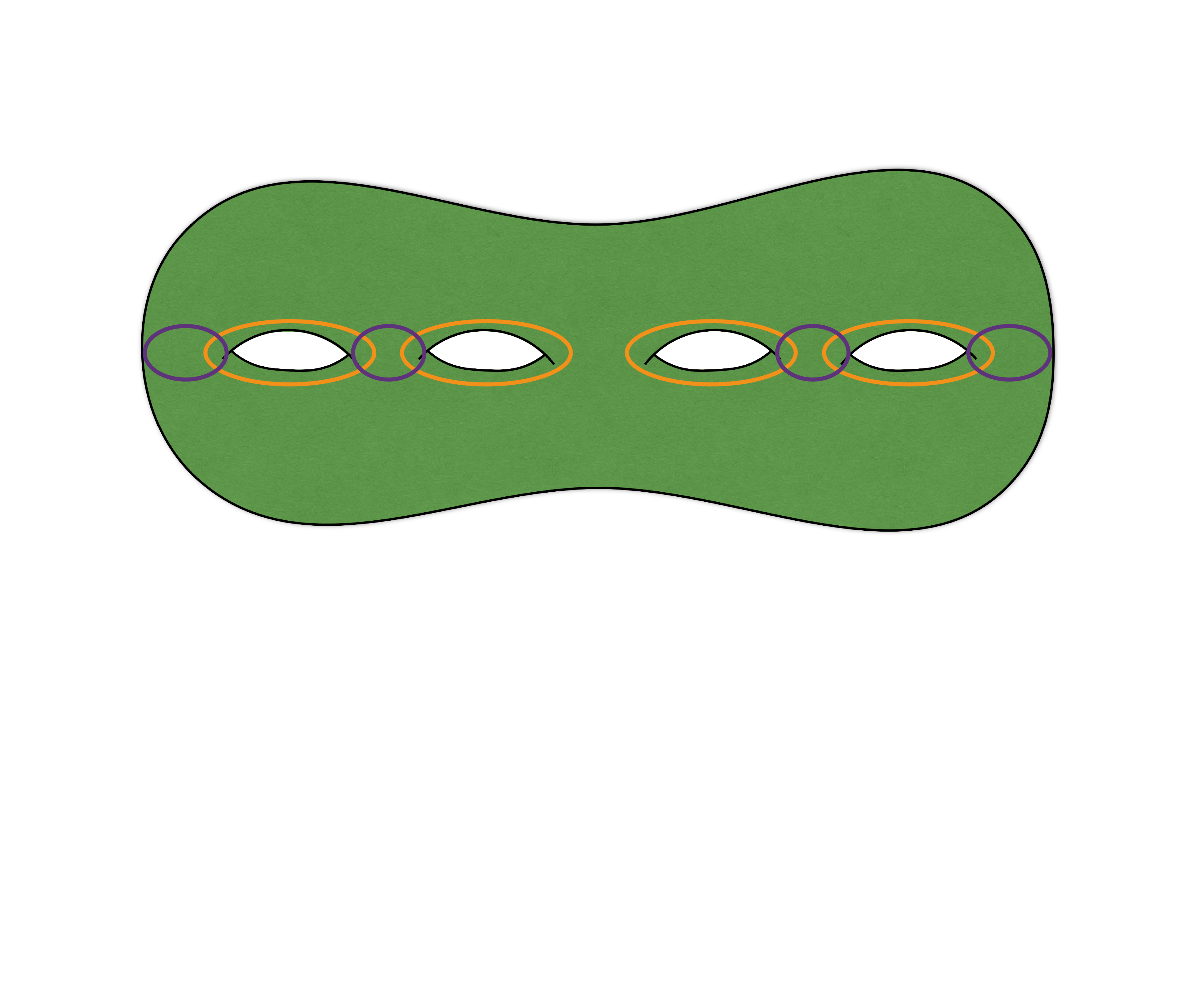}
\put(-36,41){\makebox(0,0){\footnotesize{$A$}}}
\put(-82,41){\makebox(0,0){\footnotesize{$B$}}}
\put(-133,41){\makebox(0,0){\footnotesize{$C$}}}
\put(-180,41){\makebox(0,0){\footnotesize{$D$}}}
\put(-107,79){\makebox(0,0){\footnotesize{$O$}}}
\subcaption{The alternative Riemann surface with the two choices of contractible cycles.}
\label{fig:fouraltR}
\end{subfigure}
\hfill
\begin{subfigure}{0.49\textwidth}
\includegraphics[width=\textwidth]{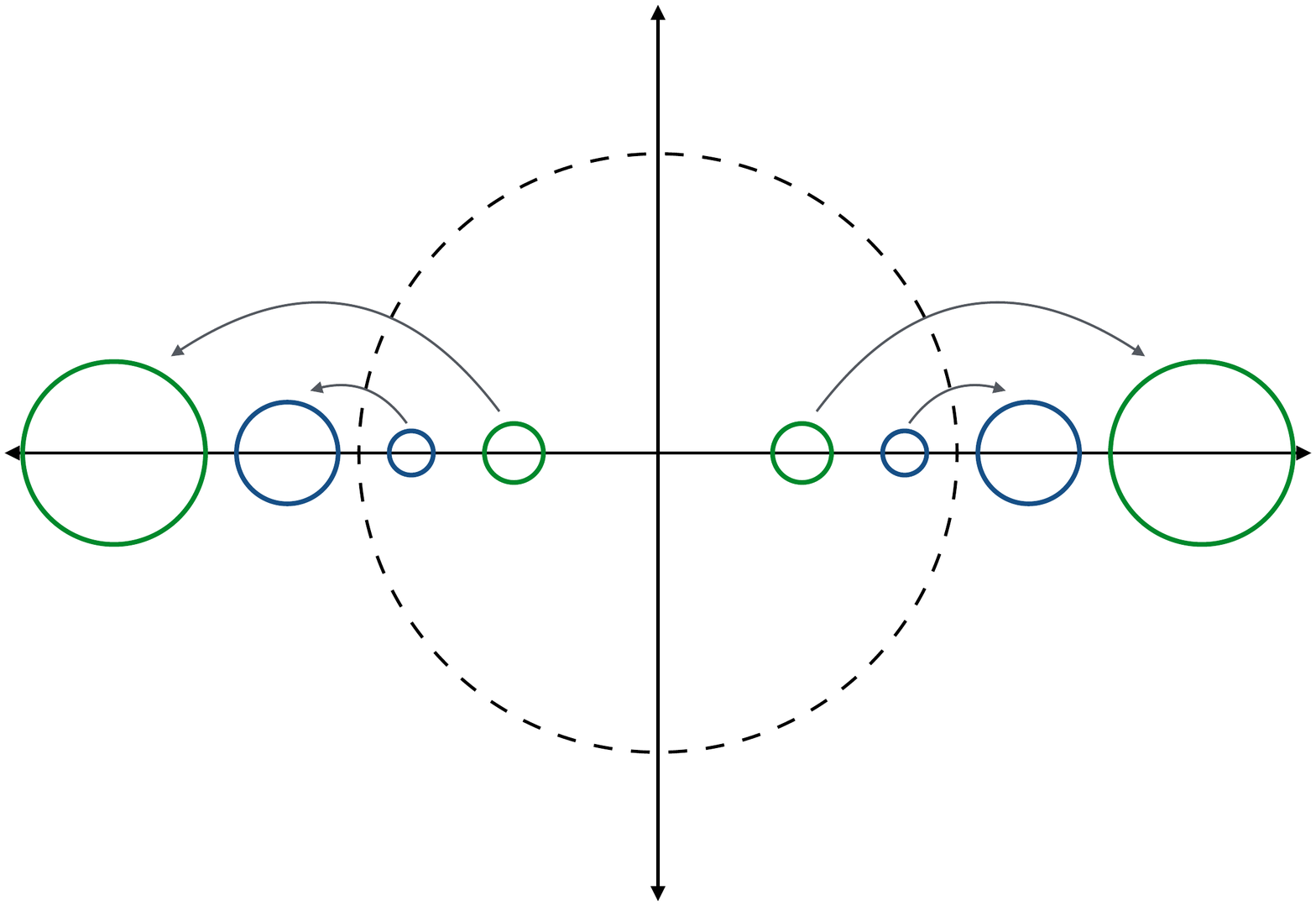}
\put(-56,96){\makebox(0,0){\tiny{$L_1$}}}
\put(-65,112){\makebox(0,0){\footnotesize{$L_2$}}}
\put(-157,98){\makebox(0,0){\tiny{$L_3$}}}
\put(-152,112){\makebox(0,0){\footnotesize{$L_4$}}}
\subcaption{The Schottky domain for both sets of cycles.}
\label{}
\end{subfigure}
\caption{In this case the two phases differ only in moduli and have the same Schottky domain.}
\label{}
\end{figure}

Note that the minimal violation of the constraints happen at the point in moduli space (dashed vertical line in Fig.~\ref{fig:resultSCDequal}) where there is an additional $\pi/2$ rotational symmetry in the Riemann surface. This observation suggests that even if we were to look outside of the two dimensional subspace of moduli space that we have considered thus far, i.e. even if we consider regions of moduli space with less symmetry, we will still be unable to make the desired extremal ray dominate our path integral.

\begin{figure}[t!]
{\includegraphics[width=0.75\textwidth]{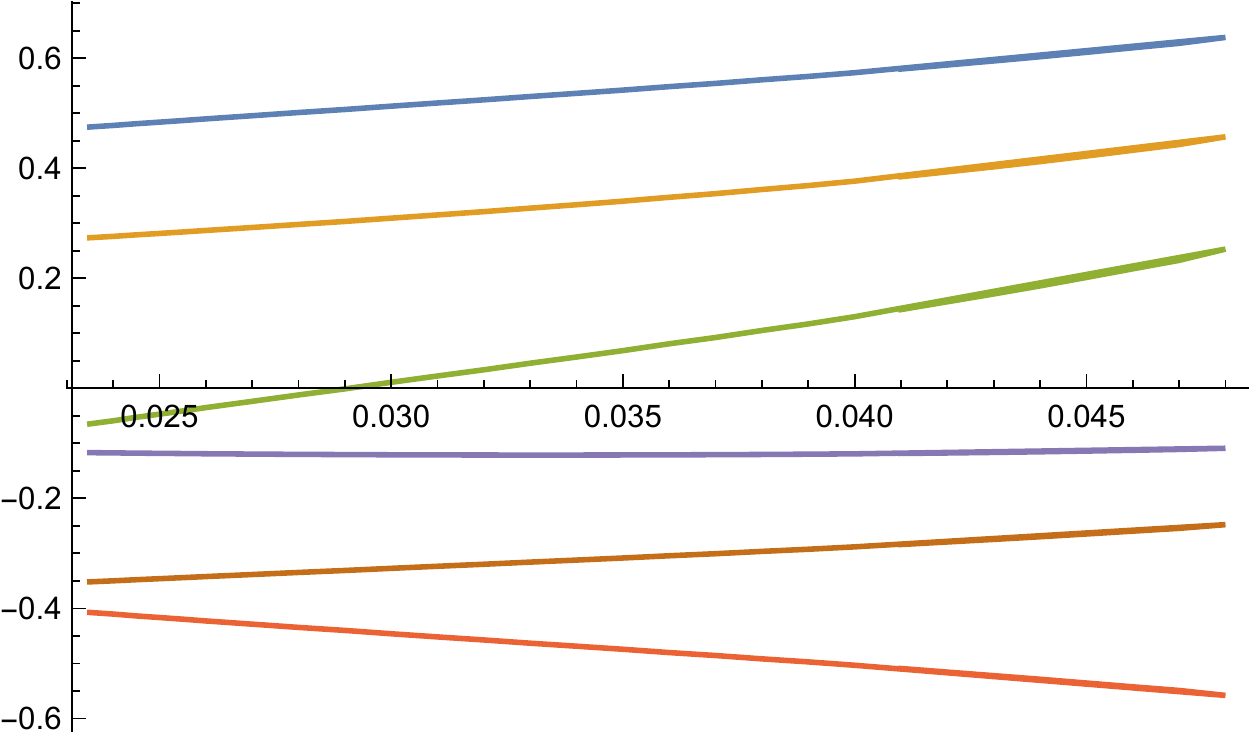}}
\centering
\hspace*{2.5cm}
\put(-30,180){\makebox(0,0){$\textcolor[rgb]{0.368417, 0.506779, 0.709798}{\blacksquare}\,\,\frac{|h_{BC}|}{2l}-1$}}
\put(-30,150){\makebox(0,0){$\textcolor[rgb]{0.880722, 0.611041, 0.142051}{\blacksquare}\,\,\frac{|h_{AC}|}{2l}-1$}}
\put(-30,120){\makebox(0,0){$\textcolor[rgb]{0.560181, 0.691569, 0.194885}{\blacksquare}\,\,\frac{|h_{AB}|}{2l}-1$}}
\put(-30,90){\makebox(0,0){$\textcolor[rgb]{0.528488, 0.470624, 0.701351}{\blacksquare}\,\,\frac{|h_{AD}|}{2l}-1$}}
\put(-30,60){\makebox(0,0){$\textcolor[rgb]{0.772079, 0.431554, 0.102387}{\blacksquare}\,\,\frac{|h_{ABC}|}{3l}-1$}}
\put(-30,30){\makebox(0,0){$\textcolor[rgb]{0.922526, 0.385626, 0.209179}{\blacksquare}\,\,\frac{|h_{ABD}|}{3l}-1$}}
\put(-75,100){\makebox(0,0){\footnotesize{$\phi$}}}
\caption{The constraints \eqref{eq:constraints2} for the alternative handlebody (Fig.~\ref{fig:fouraltR}) in the subspace where the connected and disconnected actions are equal.}
\label{fig:alt4results}
\end{figure}
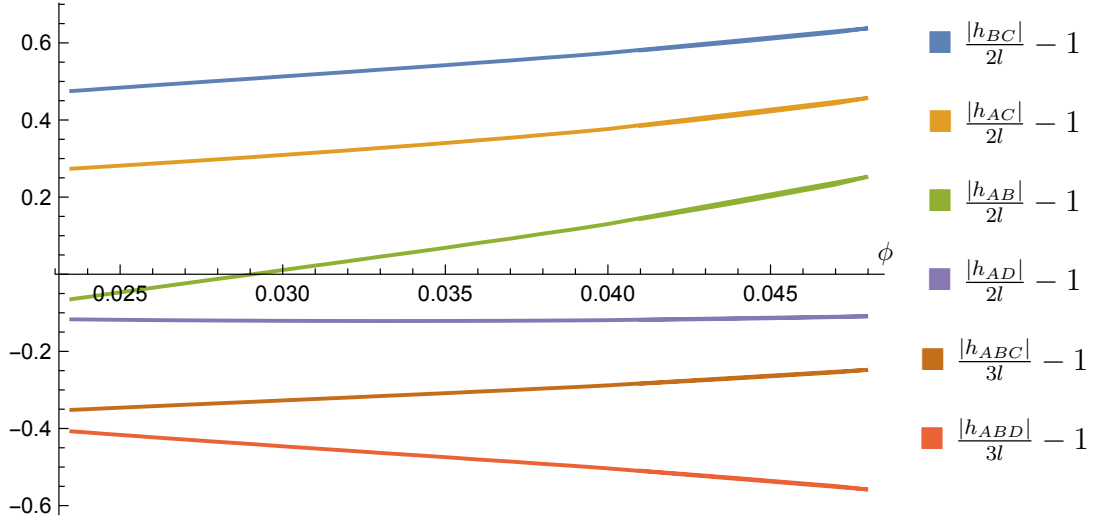

As a check we have also considered a different two dimensional subspace corresponding to the Riemann surface drawn in Fig.~\ref{fig:fouraltR}. As before, we can study the one dimensional subspace where the connected and disconnected actions are equal. The results for the various constraints on the cycle lengths are shown in Fig.~\ref{fig:alt4results}, one can see that there is no region where they are all satisfied simultaneously.

%

\subsection{Three party extremal ray}
\label{subsec:three}

\begin{figure}[bt]
\centering
{\includegraphics[width=0.5\textwidth]{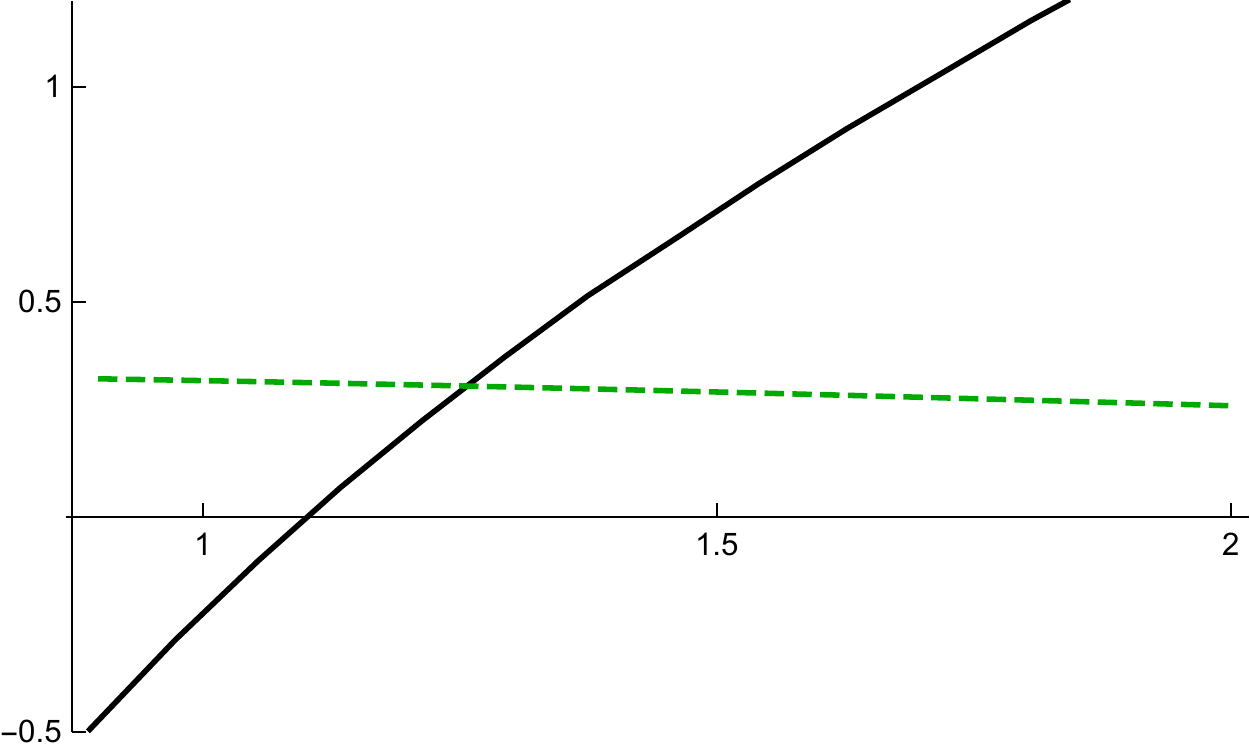}}
\put(-4,105){\makebox(10,-110){\footnotesize{$\phi$}}}
\caption{Results for the highly symmetric three genus Riemann surface. The black (solid) line is $I_{\text{con}}-I_{\text{dis}}$, the green (dashed) one $1-\frac{|h_{AB}|}{2l}$.}
\label{3results}
\end{figure}

We briefly comment on the case of $\mathcal{RTC}_3$. In this case the extremal ray of interest is
\begin{equation}
(S_A,S_B,S_C,S_{AB},S_{AC},S_{BC},S_{ABC})=(1,1,1,2,2,2,1)\,.
\end{equation}
For the one-dimensional subspace of moduli space where the Riemann surface has three holes arranged in a circle, and an extra $2\pi/3$ rotational symmetry, the results are shown in Fig.~\ref{3results}. They suggest in this case as well that there might not be a CFT state dual to the particular multiboundary geometry considered to realize the desired extremal ray. Although our argument is far from conclusive in proving that the RT cone for three parties is not polyhedral, it is interesting to contrast the result with the quantum mechanical case, where for three parties the cone is known to be polyhedral.

%


\section{Discussion}
\label{sec:discussion}

Our findings show that the argument of \citep{Bao} for the polyhedrality of the RT cones $\mathcal{RTC}_3$ and  $\mathcal{RTC}_4$ was not conclusive. While \citep{Bao} showed the polyhedrality of the metric entropy cones $\mathcal{MC}_3$ and  $\mathcal{MC}_4$, if one further requires that the geometries realizing the extremal rays of the cones are dual to CFT states it is not clear whether these rays can be realized holographically. We have also noted that, the wormhole solutions suggested in \citep{Bao} may fail to satisfy this requirements since they do not correspond to the dominant saddle points of the natural bulk Euclidean path integral in any region of moduli space. We have not ruled out the possibility that they might dominate in other more complicated path integrals, but this seems unlikely.  For $\mathcal{RTC}_N$ with larger $N$ the situation is even less clear. While \citep{Bao} proved that the metric entropy cone is polyhedral for any $N$, without the full list of RT inequalities it is obviously not possible to construct the extremal rays and explore whether they have holographic duals.

Though our results are suggestive, they are not conclusive with regard to the polyhedrality of $\mathcal{RTC}_3$ and  $\mathcal{RTC}_4$. For example, there could be other bulk solutions dual to CFT states that realize the same extremal rays. As noticed in \citep{Bao}, restricting spin structures can remove otherwise-dominant saddles and perhaps allow the desired saddle to dominate.  However, for theories with a fixed number $F$ of bulk Fermion fields this seems unlikely for large enough $N$.  For example, for $F=1$ there is no spin structure that guarantees connectivity beyond $N=2$.  If there are holographic bulk theories with arbitrarily large $F$, one could use them to show equality of the metric entropy and HRT cones.  However, it is currently unknown whether such theories exist.

Alternatively, it might be that one can in some sense subtract off the part of the CFT state corresponding to the dominant phase to leave a CFT state dual to the subdominant phase. Indeed, one naively expects that the state defined by our path integral admits a semi-classical expansion of the form
\begin{equation}
|\psi \rangle = \sum_{{ {\text{saddles} \ n} \atop {\text{corrections} \ m}}} e^{-S_n/2}c^{-m/2} |g_n,m\rangle,
\end{equation}
in terms of the central charge $c$.  Here $|g_n,m\rangle$ is a state that includes perturbative quantum corrections about the background geometry $g_n$  corresponding to the $n$th saddle. Since each geometry $g_n$ that appears in our sum is also associated with a more natural path integral to which it should contribute, one expects that a suitable linear combination of these states (acted upon by appropriate CFT operators to bring into alignment the states of the bulk quantum fields) will feature a leading term proportional to the part $|\tilde g, 0\rangle$ associated with the desired geometry (e.g. Fig.~\ref{fig:four}).

However, there are many details to be analyzed in fleshing out this more complicated approach to constructing CFT states dual to geometries like the one in Fig.~\ref{fig:four}.  One should thus be open to the possibility that it fails and that the RT cone is in fact not polyhedral. While by construction it is certainly true that the RT cone is a subset of the metric entropy cone, it could be that (at least for some $N$) it is in fact a proper subset. This would mean that there should be other entropy inequalities, possibly non-linear, that by the same argument of \citep{Bao} could not be proved by usual ``cutting and pasting'' procedures using the RT formula.

Finally, we remind the reader that all our discussion, as well as that in \citep{Bao}, was based on the RT prescription and only applies to spacetimes with a moment of time symmetry. In dynamical situations one should instead use the HRT prescription \citep{HRT}. While $\mathcal{SSA}$ and $\mathcal{MMI}$ have been proved also in this more general situations \cite{Wall:2012aa}, and by construction the metric entropy cone based on RT has to be a subset of the one based on HRT, it is not clear at present whether the polyhedrality proof of \citep{Bao} extends to the dynamical case, nor it is known if the new inequalities for $N=5$ found by \citep{Bao} still hold. We see these as interesting open questions for future investigations.


\section*{Acknowledgements}
It is a pleasure to thank Ning Bao, Veronika Hubeny, Henry Maxfield, Mukund Rangamani, Bogdan Stoica and James Sully for useful conversations. MR also wants to thank the Yukawa Institute in Kyoto for the hospitality during the final stage of this work. DM and JW were supported in part by the U.S. National Science Foundation under grant number PHY15-04541 and also by the University of California. MR was supported by the Simons Foundation via the ``It from Qubit'' collaboration and by funds from the University of California.

\appendix

\section{Bulk action computation}
\label{sec:reduction}

In this appendix we present more details for computing the Einstein-Hilbert action. As shown in \cite{MRW} we can write the bulk action as
\ban{
	I = - \frac c {24 \pi} \left[ I_\text{TZ}[\phi] - A - 4\pi(g-1)(1-\log 4R_0^2)\right]\,	,
	}
where $R_0$ is a normalization parameter corresponding to the size of the sphere on which the partition function evaluates to one. As in \citep{MRW} we set $R_0=c=1$, and so the action for a particular non-handlebody solution vanishes. Additionally, the action $I_\text{TZ}[\phi]$ is equivalent to the Takhtajan-Zograf action for a scalar field \cite{TZ}. As shown in \cite{MRW}, if we define $R_k$ to be the radius of $C_k$, and $\Delta_k$ as the distance between the center of $C_k$ and the point $w_\infty^{(k)}$ mapped to $\infty$ by $L_k$, the action can be written as
\ban{
	I_{TZ}[\phi] = \int_D d^2 w\left( \left(\nabla \phi\right)^2 + e^{2\phi} \right) + \sum_k\left( \int_{C_k} 4 \phi\,  d\theta_\infty^{(k)}  - 4 \pi \log \left |R_k^2 - \Delta_k^2 \right|\right)	 \, .\label{TZA}
}

We follow the convention of \cite{MRW} where the orientation of a $d\theta$ element associated with a particular circle is inherited from its orientation as the boundary of $D$. In practice, this means that almost all of the elements have the opposite orientations one would naively expect.
	
In all the Schottky domains we consider, there is a symmetry of the plane given by an inversion through the unit circle ($\partial U$). Therefore it will be convenient to reduce \eqref{TZA} to integrals only over the part of $D$ inside the unit circle, which we denote as $\tilde D$. As shown in \cite{MRW} we can use the transformation properties of $\phi$ to show that
\ban{
	\int_D d^2w \left(\nabla \phi\right)^2 = 2 \int_{\tilde D} d^2w \left(\nabla\phi\right)^2 + 4 \int_{\partial\tilde D}\left(\phi + \log |w| \right)  \, d\theta \, \label{eq:symred},
	}
where the coordinate $\theta$ refers to the angle measured from the origin. In practice, the boundary of $\tilde D$ will consist of a set of disjoint circles $\{\partial D_i\}= \tilde D \cap \{C_k, C_k'\}$ and the boundary of the unit disk $U$. Furthermore, we can explicitly evaluate the integral of $\log|w|$ over the boundary circles and write \eqref{eq:symred} as
\ban{
	\int_D d^2w \left(\nabla \phi\right)^2 = 2 \int_{\tilde D} d^2w \left(\nabla\phi\right)^2 + 4 \int_{\partial U} \phi\, d\theta + \sum_i \left[4 \int_{\partial  D_i} \phi \,d\theta + 4\pi \log \left( 1- \frac {R_i^2}{X_i^2}\right)\right]	\, ,
}
where $X_i$ is the distance between the center of $\partial D_i$ and the origin. In all of the domains we consider, $X_i$ is zero only for the unit and not included in the sum. Additionally it was shown in \cite{MRW} that we can integrate the $(\nabla \phi)^2$ term by parts to get on shell
\ban{
 \int_{\tilde D} d^2w \left(\nabla\phi\right)^2 = - \int_{\tilde D} d^2w \, \phi \, e^{2\phi}-\int_{\partial U} \phi \, d\theta - \sum_i \int_{\partial D_i} \phi \, d \theta_0^{(i)}  \, ,
}
where $\theta_0^{(i)}$ is the angular coordinate as measured from the center of the boundary circle $\partial D_i$. Putting everything together, we can then write
\ban{
	\int_D d^2w \left(\nabla \phi\right)^2 = -2 \int_{\tilde D} d^2w \, \phi \, e^{2\phi} + 2 \int_{\partial U} \phi\, d\theta + \sum_i \left[2 \int_{\partial D_i} \phi \,(2d\theta-d\theta_0^{(i)}) + 4\pi \log \left( 1- \frac {R_i^2}{X_i^2}\right)\right]	\, .
}

Next, we can tackle the sum over $C_k$ in the formula \eqref{TZA}. We make the assumption that neither $C_k$ nor $C_k'$ intersect the unit circle, the case in which they do was worked out in \cite{MRW}. Note that we can divide the circles into two classes, one in which one of $C_k$ or $C_k' \subset U$ and one in which $C_k, C_k ' \not\subset U$. However, with the additional assumption that there is an inversion symmetry through $U$, for every pair of circles $C_k, C_k' \not\subset U$ there is a pair of circles $C_{\bar k}, C_{\bar k}' \subset U$. Additionally, only one from the pair of circles $C_k,C_k'$ enters the above formula, and for each pair there is a freedom in choosing which one to be $C_k$ and $C_k'$. For each of these pairs we label $C_k$ to be the one inside $U$. Therefore we can write the sum of integrals over $C_k$ as follows:
\ban{
\sum_k4 \int_{C_k}  \phi\,  d\theta_\infty^{(k)}  = \sum_{k :\, C_k'\not\subset U}4 \int_{C_k} \phi\,  d\theta_\infty^{(k)} + \sum_{k :\, C_k'\subset U}4 \left(\int_{C_k}  \phi\,  d\theta_\infty^{(k)} +\int_{C_{\bar k}}  \phi\,  d\theta_\infty^{(\bar k)} \right) \, .
}
Of the three terms, only the last one involves an integral over a circle not in $U$, but using the inversion symmetry we can reduce it to an integral over $C_k$
\ban{
\sum_k  4\int_{C_k}  \phi\,  d\theta_\infty^{(k)}  = \sum_{k :\, C_k'\not\subset U} 4\int_{C_k} \phi\,  d\theta_\infty^{(k)} + \sum_{k :\, C_k'\subset U}4 \int_{C_k}\left(  \phi\,   +(\phi + 2\log|w|) \frac{d\theta_\infty^{(\bar k)}}{d \theta_\infty^{(k)}} \right) d\theta_\infty^{(k)}\, .
}
As the Jacobian factor and the parameters $R_k$ and $D_k$ can all be computed analytically from the setup of the Schottky domain, the only numeric integration occurs within the domain $\tilde D$.

Finally, we note that $A= \int_D e^{2\phi} d^2w$ and so this integration in $I_\text{TZ}$ cancels the $A$ term in the action. Putting it all together, we have the following on shell action
\ban{
I=-\frac c {24\pi} &\left[-4\pi(g-1) \log 4 -2 \int_{\tilde D} d^2w \, \phi \, e^{2\phi} + 2 \int_{\partial U} \phi\, d\theta \right. \notag \\
&+   \sum_{k: \, C_k' \not\subset U}\left( 2 \int_{C_k} \phi \,(2d\theta+2d\theta_\infty^{(k)}-d\theta_0^{(k)}) + 4\pi \log \frac{ 1-  {R_k^2}/{X_k^2}}{\left |R_k^2 - \Delta_k^2 \right| }\right) \notag \\
&+   \sum_{k: \, C_k' \subset U}\left( 2 \int_{C_k} \phi \,(2d\theta+2d\theta_\infty^{(k)}-d\theta_0^{(k)}) +4 \int_{C_k}(\phi + 2\log|w|) \frac{d\theta_\infty^{(\bar k)}}{d \theta_\infty^{(k)}}  d\theta_\infty^{(k)} \right. \notag  \\
&\hspace{2cm} \left. \left. +2 \int_{C_{k'}} \phi \,(2d\theta-d\theta_0^{(k')})+ 4\pi \log \frac{ (1-  {R_k^2}/{X_k^2})(1-  {R_{k'}^2}/{X_{k'}^2})}{\left |R_k^2 - \Delta_k^2 \right|\left |R_{\bar k}^2 - \Delta_{\bar k}^2 \right| }\right) \right]\,.
}
The action written explicitly in this way is very useful because it only involves quantities one can compute analytically and numeric integrations over $\tilde D$ and $\partial \tilde D$.

In many scenarios one can derive relations such as $d\theta + d\theta_\infty^{(k)} = d\theta_0^{(k)}$ which greatly simplify the above formula. Additionally, one is often able to use reflection symmetries to further reduce the domain of integration. Practically, it is only numerically feasible to compute integrals over $d\theta_0^{(k)}$ and so one can introduce more Jacobian factors to convert all integrals to this form.

\bibliographystyle{jhep}

\bibliography{references}

\end{document}